\documentclass[11pt,a4paper]{article}
\pdfoutput=1

\usepackage[british]{babel} 

\usepackage{amsmath,amssymb,amsthm,mathrsfs}

\usepackage{graphicx}
\usepackage[usenames,dvipsnames]{color}
\usepackage[textwidth = 440 pt, textheight = 650 pt]{geometry}
\usepackage{setspace}
\usepackage[linktocpage=true,hypertexnames=false]{hyperref}

\usepackage[bulletsep]{collref}

\newcommand{\tb}[1]{\textbf{#1}}

\title{}
\author{}

\begin{document}
\pagestyle{empty}
\begin{center}
{\scriptsize
\texttt{CNRS-16/03, }
\scriptsize
\texttt{DCPT-16/19, }
\scriptsize
\texttt{DESY 16-083, }
\scriptsize
\texttt{DMUS-MP-16/09, }
\scriptsize
\texttt{HU-EP-16/13, }
\scriptsize
\texttt{HU-MATH-16/08, }
\scriptsize
\texttt{NORDITA-2016-33}}
\end{center}

\begin{center}
{\LARGE\tb{\mathversion{bold}
An Integrability Primer for the Gauge-Gravity Correspondence: an Introduction}\par}
\vspace{15mm}

{  D.~Bombardelli$^{a,b}$, A.~Cagnazzo$^{c,d}$, R.~Frassek$^e$, F.~Levkovich-Maslyuk$^f$, F.~Loebbert$^g$, S.~Negro$^h$, I.~M.~Sz\'{e}cs\'{e}nyi$^e$, A.~Sfondrini$^i$, S.~J.~van~Tongeren$^{g,j}$, A.~Torrielli$^k$
}\\[10mm]

\tb{Abstract}\\[5mm]
\begin{minipage}{.8\columnwidth}
We introduce a series of articles reviewing various aspects of integrable models relevant to the AdS/CFT correspondence. Topics covered in these reviews are: classical integrability, Yangian symmetry, factorized scattering, the Bethe ansatz, the thermodynamic Bethe ansatz, and integrable structures in (conformal) quantum field theory.
In the present article we highlight how these concepts have found application in AdS/CFT, and provide a brief overview of the material contained in this series.
\end{minipage}
\end{center}

\vspace{2cm}
{\scriptsize\it\noindent
{}$^a$ Dipartimento di Fisica and INFN, Universit\`a di Torino
via Pietro Giuria 1, 10125 Torino, Italy\\[-0.2cm]
{}$^b$ Dipartimento di Fisica e Astronomia and INFN, Universit\`a di Bologna,
via Irnerio 46, 40126 Bologna, Italy\\[-0.2cm]
{}$^c$
DESY Hamburg, Theory Group,
Notkestrasse 85, D-22607 Hamburg, Germany\\[-0.2cm]
{}$^d$ Nordita, KTH Royal Institute of Technology and Stockholm University, Roslagstullsbacken 23, SE-10691 Stockholm\\[-0.2cm]
{}$^e$ Department of Mathematical Sciences, Durham University, South Road, Durham DH1 3LE, United Kingdom\\[-0.2cm]
{}$^f$ Mathematics Department, King's College London,
The Strand, WC2R 2LS London, United Kingdom\\[-0.2cm]
{}$^g$ Institut f\"ur Physik, Humboldt-Universit\"at zu Berlin,
Zum Gro\ss en Windkanal 6, 12489 Berlin, Germany\\[-0.2cm]
{}$^h$ LPTENS, Ecole Normale Sup\'{e}rieure,
PSL Research University, Sorbonne Universit\'{e}s,
UPMC Univ. Paris 06,\\[-0.2cm]
CNRS UMR 8549, 24 Rue Lhomond, 75005 Paris, France\\[-0.2cm]
{}$^i$ Institut f\"ur Theoretische Physik, ETH Z\"urich, Wolfgang-Pauli-Str. 27, 8093 Z\"urich, Switzerland\\[-0.2cm]
{}$^j$ Institut f\"ur Mathematik, Humboldt-Universit\"at zu Berlin,
Zum Gro\ss en Windkanal 6, 12489 Berlin, Germany\\[-0.2cm]
{}$^k$ Department of Mathematics, University of Surrey,
Guildford GU2 7XH, United Kingdom}

{\let\thefootnote\relax\footnotetext{%
\noindent
\hspace{-0.62cm}\textit{E-mails}: {diegobombardelli@gmail.com,}
{alessandra.cagnazzo@desy.de,}
{rouven.frassek@durham.ac.uk,}
{fe\-dor.levkovich@gmail.com,}
{loebbert@physik.hu-berlin.de,}
{stefano.negro@lpt.ens.fr,}
{sfondria@itp.phys.ethz.ch,}
{i.m.szecsenyi@durham.ac.uk,}
{svantongeren@physik.hu-berlin.de,}
{a.torrielli@surrey.ac.uk}
}}
\newpage
\pagestyle{plain}

\noindent
In this article we introduce a series of articles reviewing aspects of integrable models. The articles provide a pedagogical introduction to the topic of integrability, with special emphasis on methods relevant in the AdS/CFT correspondence. After a brief motivation regarding the value of general integrable models in the development of theoretical physics, here we discuss the application of the framework of integrability to the AdS/CFT correspondence. We then provide an overview of the material contained in the various reviews, referring back to AdS/CFT applications, and indicating links between the reviews themselves and to the relevant literature. While written with an AdS/CFT background in mind, the methods covered in the reviews themselves have applications throughout the wider field of integrability.

\section*{Integrability}

Integrable models appear throughout theoretical physics, starting from classical mechanics where models such as the Kepler problem can be solved---in the sense of the Liouville theorem---by integration. In general, integrable models show special behaviour due to many underlying symmetries, symmetries due to which they can often be exactly solved. Only a fraction of the physical systems appearing in nature can be described in these terms. Nevertheless, integrable models offer insight into real-world situations through universality, or when used as a theoretical laboratory to develop new ideas. In statistical mechanics for example, many subtleties of the thermodynamic limit have been understood by working out specific models, notably phase transitions in the Lenz-Ising model and the role of boundary conditions in the ice model. In hydrodynamics, the Korteweg-de Vries equation illustrates how a non-linear partial differential equation can admit stable, wave-like localized solutions: \emph{solitons}. In condensed matter physics, both integrable quantum spin chains and one-dimensional gases of almost-free particles play a pivotal role. Finally, in quantum field theories (QFTs) in two space-time dimensions, exactly solvable models helped unravel phenomena like dimensional transmutation, as in the case of the chiral Gross-Neveu model, or concepts like bosonisation, as in the case of the sine-Gordon and Thirring models. The general framework to study such integrable QFTs, mainly associated to \emph{inverse} and \emph{factorized scattering}, was laid down in the 1970s and has found numerous applications since.

\section*{Integrability in AdS/CFT}

In recent years, the general framework of integrability has been successfully applied in the context of the AdS/CFT correspondence~\cite{Maldacena:1997re,Witten:1998qj,Gubser:1998bc}, a concrete realisation of the holographic principle~\cite{'tHooft:1993gx,Susskind:1994vu}.
According to this correspondence, string theory on anti-de Sitter (AdS) backgrounds is dual (equivalent) to conformal quantum field theory (CFT) on the ``boundary'' of AdS. The canonical example of this duality is the correspondence between closed type IIB superstrings on $\mathrm{AdS}_5\times  \mathrm{S}^5$ and $\mathcal{N} = 4$ supersymmetric Yang-Mills (SYM) theory in four dimensions. Both sides of this duality can be studied using integrability-based techniques, at least in the so-called planar limit. Similar ideas apply to lower dimensional AdS backgrounds, as well as to deformations of these backgrounds. Part of this progress is reviewed in e.g.~\cite{Beisert:2004ry,Arutyunov:2009ga,Gromov:2009zza,Serban:2010sr, Beisert:2010jr,Sfondrini:2014via,vanTongeren:2013gva}.

In the AdS/CFT context, integrability enters naturally on the string theory side as a property of particular two dimensional field theories.\footnote{Our exposition is not chronological. Some historical aspects are discussed in~\cite{Arutyunov:2009ga}.} The details of two dimensional field theories can be such that their classical equations of motion can be tackled by the \emph{inverse scattering method}, an approach initiated in the 1960s~\cite{Gardner:1967wc} and mainly developed
in the following decade~\cite{Faddeev:1971xd,Zakharov:part1,Zakharov:1979zz}. In particular, the equations of motion of such integrable field theories can be represented as the flatness of a so-called \emph{Lax connection}. Now in the planar limit, closed string theory reduces to field theory on a two-dimensional cylinder---the worldsheet of a single string---and this field theory is integrable~\cite{Bena:2003wd} in the above sense. By expanding the machinery of classical integrability it is possible to tackle the semi-classical spectrum of integrable field theories as well, resulting in what is known as \emph{finite-gap equations}~\cite{Novikov:1982ei}. The semi-classical limit of our closed string can be approached in this spirit~\cite{Kazakov:2004qf,Beisert:2005bm}, see also~\cite{Vicedo:2011zz}. Moving beyond the semi-classical spectrum is more involved, as will come back shortly.

The integrability appearing on the CFT side of AdS/CFT is that of integrable spin chains. Integrable spin chains such as the Heisenberg spin chain can be solved by the \emph{Bethe ansatz} \cite{Bethe:1931hc}. This is an ansatz for the eigenfunctions of a spin-chain Hamiltonian, written in terms of collective excitations called magnons or spin waves, and their scattering matrix (S matrix). There is an underlying algebraic structure however, based on an \emph{$R$ matrix} and \emph{Lax matrix}, which leads to the algebraic Bethe ansatz also known as the \emph{quantum inverse scattering method} \cite{Korepin:1993book,Faddeev:1996iy}. Historically, the appearance of an integrable spin chain was the first indication of integrability in AdS/CFT \cite{Minahan:2002ve}.\footnote{In the context of gauge theory, integrability was previously encountered in high energy hadron scattering in QCD \cite{Lipatov:1993yb,Faddeev:1994zg}.} Working in $\mathcal{N} = 4$ SYM theory at one loop order, Minahan and Zarembo showed \cite{Minahan:2002ve} that by identifying single-trace operators with particular spin-chain states, the dilatation operator---whose eigenvalues yield the anomalous dimensions of such operators---becomes the Hamiltonian of an integrable spin chain \cite{Beisert:2003jj,Beisert:2003yb}, whose spectrum can be found via the Bethe ansatz.

Though different in their appearance, these two types of integrability share a common symmetry structure. Two-dimensional integrable field theories typically have infinitely many conserved charges that can be packaged into a powerful algebraic structures, and these same structures come back in spin chains. A prototypical example of such a structure, particularly important in AdS/CFT, is the Yangian algebra~\cite{Luscher:1977uq,Bernard:1990jw}. Indeed, the classical integrability of the string brings with it an infinite set of conserved charges~\cite{Bena:2003wd}, which form a Yangian algebra that can also be seen as the symmetry of the quantum spin chain of SYM \cite{Dolan:2003uh}.

Going beyond one loop in SYM, or semi-classics in string theory, requires more work, but is possible. The presence of integrable structures at higher loops in SYM~\cite{Beisert:2003ys,Beisert:2003tq,Beisert:2004hm,Beisert:2005fw,Beisert:2005tm, Beisert:2006ez} makes it possible to find an exact S matrix for the spin chain magnons~\cite{Beisert:2005tm, Beisert:2006ez} and write down an \emph{asymptotic Bethe ansatz}~\cite{Beisert:2005fw}. This S~matrix and Bethe ansatz have counterparts in the dual string theory~\cite{Arutyunov:2004vx,Hernandez:2006tk,Janik:2006dc, Beisert:2006ib, Arutyunov:2006ak,Arutyunov:2006yd}. There, the S matrix is simply the worldsheet S~matrix of the light-cone gauge-fixed string~\cite{Arutyunov:2006ak,Arutyunov:2006yd}. To define this S~matrix and the associated asymptotic states we need to take the limit where the length of the gauge-fixed string (volume of the theory) goes to infinity. The asymptotic Bethe ansatz then arises by re-imposing periodic boundary conditions on approximate wavefunctions obtained from the S matrix using the ideas of \emph{factorized scattering}.

The reason for distinguishing the asymptotic Bethe ansatz from the (exact) Bethe ansatz is clear on the string theory side: merely imposing periodic boundary conditions while working with the S matrix of the infinite length string, neglects possible virtual particles wrapping around the worldsheet (cylinder)~\cite{Luscher:1985dn,Luscher:1986pf,Bajnok:2008bm}. This failure of the asymptotic Bethe ansatz for the string~\cite{Ambjorn:2005wa} is paralleled by a similar breakdown in the $\mathcal{N}=4$ SYM spin-chain~ \cite{Sieg:2005kd}. Here the dilatation operator features interactions whose range increases with the loop order, so that eventually the interaction range is of the order of the length of the composite operator under consideration, and the Bethe ansatz breaks down. In relativistic field theory models these \emph{finite size effects} can be understood by integrability techniques \cite{Zamolodchikov:1989cf,Dorey:1996re,Bazhanov:1996aq} using the \emph{thermodynamic Bethe ansatz}~\cite{Yang:1968rm}. Extending these ideas to the integrable string sigma model gives the AdS${}_5$/CFT${}_4$ thermodynamic Bethe ansatz~\cite{Arutyunov:2009zu,Arutyunov:2009ur,Bombardelli:2009ns,Gromov:2009bc} and its improvement known as the \emph{quantum spectral curve}~\cite{Gromov:2013pga}. It is now possible to compute the energy of closed string states nonperturbatively with arbitrary numerical precision, or analytically in a weak coupling expansion up to loop orders prohibitively difficult to reach by conventional techniques.

The chain of reasoning leading up to this description of the spectral problem involves various unproven albeit well-tested assumptions, in particular the hypothesis that integrability persists at the quantum level at arbitrary coupling. In some simpler models---specific conformal field theories and their massive deformations---the resulting structures can be more rigorously derived from first principles by methods introduced by Baxter~\cite{Baxter:1972hz} and developed by Bazhanov, Lukyanov and Zamolodchikov \cite{Bazhanov:1994ft}. Doing so in the AdS/CFT context would undoubtedly provide remarkable insights, but thus far the answer appears to be elusive \cite{Benichou:2011ch,Delduc:2012vq}.

\section*{The articles}

Beyond the spectral problem for AdS$_5$/CFT$_4$ described above, integrability based approaches to other observables and other instances of AdS/CFT are being actively pursued. While this landscape is motivation for this series of articles, we do not aim to review all of it. Rather, we review some of the key ideas upon which the progress in this field is based, ideas which can often be introduced and understood in simpler models. In fact, these ideas and methods are central to many integrable models and not just those appearing in AdS/CFT, and as such the material presented in this series is relevant to integrability in general.

Our aims have required us to make choices: we will cover a relatively broad set of techniques and highlight how they are related to each other, at the expense of the details of the many models where they can be applied. Our key example is the chiral Gross-Neveu model, as a good compromise between keeping relevant features of general integrable models and reducing technical complications. Where appropriate, the individual chapters contain further references to the AdS/CFT or integrability literature.

Below we give a detailed overview of each of the chapters, appearing in their suggested reading order~\cite{Torrielli:2016ufi, Loebbert:2016cdm, Bombardelli:2016scq, Levkovich-Maslyuk:2016kfv, VanTongeren:2016hhc, Negro:2016yuu}. The lectures on Classical Integrability and Yangian Symmetry give the historical and mathematical background for the other lectures. Afterwards we turn our attention to scattering matrices, with special focus on integrable scattering in two dimensional QFTs. Building on this, we discuss how to obtain (asymptotic) Bethe ansatz equations, both in the original and algebraic approach. Next, we move to the thermodynamic Bethe ansatz as a tool to describe integrable models either at finite temperature or at finite size. The last article explores the relation between the symmetries stemming from integrability and those of conformal symmetry in two-dimensional QFTs, tying together most of the material presented in the previous articles. We have aimed to keep notation uniform throughout the articles.

\paragraph{Chapter I: Classical Integrability.}
The chapter on Classical Integrability~\cite{Torrielli:2016ufi} deals with classical Hamiltonian systems which are integrable by Liouville's theorem, and explores the algebraic techniques which are available to exactly solve such systems. This part is mostly concerned with the classical inverse scattering method, where Lax pairs and r-matrices are treated and their properties outlined, culminating in a discussion of soliton solutions and the Gel'fand-Levitan-Marchenko equation. Although most of the these topics are reviewed in standard monographs, such as \cite{Babelon,Faddeev:1987ph,Novikov:1984id}, some of the algebraic aspects---such as the Belavin-Drinfeld theorems \cite{BelavinDrinfeld1}---tend not to be, and are here presented in a compact uniform fashion.

This section introduces tools of classical integrability that play an important role in describing strings on various anti-de~Sitter spaces \cite{Arutyunov:2009ga, Beisert:2010jr}. Although the applications to string theory are rich of algebraic complications (see for instance \cite{Tseytlin:2010jv,Magro:2010jx,Sfondrini:2014via}), the basic ideas are practically the same as those contained in this review, which therefore works as a good entry point for anyone interested in delving into the modern topics connected to AdS/CFT integrability in the strong coupling (classical string theory) regime.

\paragraph{Chapter II: Yangian Symmetry.}

The quantum Yang-Baxter equation represents one of the most prominent features of integrable models.
The lectures on Yangian symmetry of this chapter~\cite{Loebbert:2016cdm} deal with the algebraic structure that underlies rational solutions to this equation. The Yangian beautifully extends the concepts of classical integrability reviewed in Chapter~I \cite{Torrielli:2016ufi}. Mathematically, this symmetry enhances an ordinary Lie algebra to a so-called \textit{quantum group} with the structure of a Hopf algebra.\footnote{The Yangian represents one member of the family of quantum groups that are found in integrable models.}
In physical models, the crucial difference of the Yangian to ordinary Lie algebra symmetries lies in the fact that the Yangian generators represent \textit{non-local} symmetries, which act on a discrete or continuous space.
This one-dimensional space can be realized in many different ways making the Yangian a rather universal concept with strong implications for a given theory---classical or quantum.

In particular, the Yangian appears in the context of (1$+$1)-dimensional field theories, in spin chain models and it underlies the integrability of the AdS/CFT correspondence. Its prime application is to bootstrap integrable scattering matrices which are discussed in Chapter~III \cite{Bombardelli:2016scq}, and its algebraic structure provides the basis for the Bethe ansatz reviewed in Chapter~IV \cite{Levkovich-Maslyuk:2016kfv}. Keeping an eye on the historic development, we provide an introduction to the subject that contains both the more mature discussions of Yangian symmetry in two-dimensional models (see e.g.\ \cite{Bernard:1992ya, MacKay:2004tc}), as well as its modern application to the gauge/gravity duality (see e.g.\ \cite{Beisert:2010jq,Bargheer:2011mm,Torrielli:2011gg,Spill:2012qe}).
Generic definitions and concepts are illustrated by means of examples including the two-dimensional chiral Gross-Neveu model, the Heisenberg spin chain, as well as $\mathcal{N}=4$ super Yang-Mills theory in four dimensions.
These lectures aim at providing an introductory overview, which draws connections between different physical applications and mathematical aspects of the rich subject of Yangian symmetry.

\paragraph{Chapter III: S~matrices and Integrability.}
The third chapter \cite{Bombardelli:2016scq} of this collection deals with a fundamental object in quantum integrable theories: the S matrix, {\it i.e.} the operator that maps initial to final states in a scattering process.

First of all, knowing the S matrix is crucial for calculating the energy spectra in the large volume limit, via the derivation of the asymptotic Bethe ansatz, as will be reviewed in Chapter IV \cite{Levkovich-Maslyuk:2016kfv}.
Beyond the asymptotic regime, the S matrix is a key ingredient for the leading and exact finite-size corrections of the energies, calculated by the L\"uscher formulas and the thermodynamic Bethe ansatz respectively, both discussed in Chapter~V \cite{VanTongeren:2016hhc}.

The miracle happening in (1+1)-dimensional integrable models, as we will explain in this chapter~\cite{Bombardelli:2016scq}, is the possibility to determine the S matrix exactly, due to the highly constraining conservation laws of these
particular theories and few analytical assumptions. This also makes it possible to derive the S~matrices for bound states, if any, of the theory.

From an algebraic point of view, the S matrix can often be identified with a representation of the universal R-matrix of a Hopf algebra, reviewed in Chapter~II \cite{Loebbert:2016cdm}.
This places the properties of the S matrix in an algebraic setting, and allow us to generalize its derivation also beyond the relativistic case.

Finally, the role played by the S matrix in the determination of form factors will also be briefly mentioned, and the S~matrices of sine-Gordon, $\mathrm{SU}(2)$ and $\mathrm{SU}(3)$ chiral Gross-Neveu models will be discussed.
Through these examples, it will be possible to show how to derive the exact S~matrices and some simple form factors in practice, both for fundamental and bound states. We will also briefly discuss non-relativistic S~matrices and overview their applications to the AdS/CFT correspondence.

\paragraph{Chapter IV: The Bethe Ansatz.}
Bethe ansatz techniques originated from the exploration of spin chains as models of condensed matter systems.
The same methods also turned out to play a key role in computing the spectrum of 2d integrable field theories. These two applications have been recently united in the context of integrability in the AdS/CFT correspondence. The Bethe ansatz in AdS/CFT \cite{Beisert:2005fw} realizes a beautiful interpolation between integrable spin chains on the gauge theory side \cite{Minahan:2002ve} and the integrable structure of a 2d sigma model on the string theory side \cite{Arutyunov:2004vx}.

Chapter~IV of this collection \cite{Levkovich-Maslyuk:2016kfv} covers various aspects of the Bethe ansatz in a pedagogical way, serving as a preparation for understanding its applications in AdS/CFT. This chapter logically continues the article dedicated to exact S~matrices \cite{Bombardelli:2016scq}. It is discussed how, knowing the S~matrix, we can use the Bethe ansatz to find the theory's \textit{non-perturbative} spectrum, albeit only in large volume. As explicit examples, the two-dimensional $\mathrm{SU}(2)$ and $\mathrm{SU}(3)$ chiral Gross-Neveu models are considered. We will see that to compute the spectrum one should first solve an auxiliary spin chain, which in these cases is the famous Heisenberg XXX model. Its solution is covered in detail, including the
coordinate and the algebraic Bethe ansatz approaches, as well as the nested Bethe ansatz in the $\mathrm{SU}(3)$ case. It is also demonstrated that in the classical limit the Bethe equations encode a Riemann surface known as the spectral curve of the model. Finally, it is shown how the familiar 1d oscillator in quantum mechanics can be solved via a Bethe ansatz-like method.

\paragraph{Chapter V: The Thermodynamic Bethe Ansatz.} The thermodynamic Bethe ansatz (TBA) is a method used to describe the thermodynamics of integrable systems solved by the Bethe ansatz, resulting in a set of integral equations whose solution determines the free energy of the model in thermodynamic equilibrium. After its inception at the end of the sixties by Yang and Yang \cite{Yang:1968rm} to describe the thermodynamics of the one dimensional Bose gas with delta function interaction (the Lieb-Liniger model), the TBA was quickly and broadly adopted. Its use now ranges from describing the thermodynamics of integrable spin chain models such as the XXZ spin chain \cite{Gaudin:1971gt,Takahashi:1972}, to computing the spectra of integrable field theories on circles of finite circumference \cite{Zamolodchikov:1989cf,Dorey:1996re,Bazhanov:1996aq} and beyond. This is how the TBA originally entered in the AdS/CFT correspondence for instance: the exact energy spectrum of the $\mathrm{AdS}_5\times \mathrm{S}^5$ superstring is encoded in a set of TBA equations \cite{Arutyunov:2009zu,Arutyunov:2009ur,Bombardelli:2009ns,Gromov:2009bc}.  At the same time, equations of TBA type arise in determining the area of classical string worldsheets \cite{Alday:2009dv} for example.

The fifth chapter of this series \cite{VanTongeren:2016hhc} provides an introduction to the TBA, focussing on the conceptual ingredients---root distributions, counting functions, particle and hole densities, the string hypothesis in case of bound states---that underlie this method. We illustrate this discussion on concrete examples, starting from simple free electrons, then the original Bose gas, and finally the XXX spin chain and $\mathrm{SU}(2)$ chiral Gross-Neveu model as respectively spin chain and field theory examples with nontrivial string hypotheses. We also discuss the simplification of TBA equations, the derivation of Y systems from TBA equations and the equivalence between the two modulo analyticity data, and the use of the TBA in finite volume integrable field theory, including excited states and L\"uscher formulae.

\paragraph{Chapter VI: Integrable Structures in Quantum Field Theory.} The expression ``integrable structures'' appearing in the title of this article can be interpreted in two different ways. On the one hand, it is used as a label for fundamental objects appearing in quantum integrable models, that is to say integrals of motion, transfer matrices, Baxter $Q$-operators and so on. There exists, however, a broader meaning to this expression, referring to the nature and the properties of the algebraic foundations on which the quantum integrable theories stand. The fundamental objects named above then appear as the main characters in the story of the integrable structures. This tale has been known for decades in the case of spin chains and lattice models~\cite{Baxter:1972hz}, but it was only in the nineties that it was first told for a 2D quantum field theory \cite{Bazhanov:1994ft,Bazhanov:1996dr,Bazhanov:1998dq}. The approach of Bazhanov, Lukyanov and Zamolodchikov, nowadays referred to as the BLZ method, was the first successful attempt at the construction of the fundamental integrability objects from the algebraic structure of a field theory. Although this does not deal directly with theories associated to sigma models and AdS/CFT, it nonetheless provides general recipes with broader applications \cite{Bazhanov:1996aq,Bazhanov:2001xm,Bazhanov:2013cua}. Another important, pedagogical aspect, is that the BLZ method employs an array of mathematical concepts with connections to most approaches to integrability. In this way, the sixth chapter of this series \cite{Negro:2016yuu} can serve as a playground where the methods and concepts discussed in the other chapters can be put in motion.

\section*{Acknowledgements}

This is the introduction to a series of articles that arose out of lectures presented at the GATIS \textit{Young Researchers Integrability school} organized at Durham University, 6-10 July 2015. We would like to thank the scientific committee and the participants for their input and feedback and the Department of Mathematical Sciences at Durham University for its hospitality. We are especially grateful to the GATIS network and to Patrick Dorey in particular for their support towards this initiative.
AC acknowledges support from the Foundation Blanceflor Boncompagni-Ludovisi, n\'ee Bildt.  The work of DB has been partially funded by the INFN grants GAST and FTECP, and the research grant UniTo-SanPaolo Nr TO-Call3-2012-0088 \textit{Modern Applications of String Theory} (MAST). AS's research was partially supported by the NCCR SwissMAP, funded by the Swiss National Science Foundation. AT thanks the EPSRC for funding under the First Grant project EP/K014412/1 \textit{Exotic quantum groups, Lie superalgebras and integrable systems}, and the STFC for support under the Consolidated Grant project nr. ST/L000490/1 \textit{Fundamental Implications of Fields, Strings and Gravity}. The work of SN was supported by the European Research Council (Programme “Ideas” ERC- 2012-AdG 320769 AdS-CFT-solvable). ST is supported by LT. The work of ST is supported by the Einstein Foundation Berlin in the framework of the research project \textit{Gravitation and High Energy Physics}. AC, RF, FLM, AS, IS and ST acknowledge support from the People Programme (Marie Curie Actions) of the European Union's Seventh Framework Programme FP7/2007-2013/ under REA Grant Agreement No~317089 (GATIS). No data beyond those presented in this work are needed to validate this study.

\bibliography{refs}
\bibliographystyle{nb}
\end{document}